\documentclass[aps,twocolumn]{revtex4-2}

\usepackage{amsmath,amssymb}
\usepackage{mathrsfs}
\usepackage{graphics,graphicx}
\usepackage{dcolumn,bm}
\usepackage{breqn}
\usepackage{psfrag}
\usepackage{color}
\usepackage{appendix}
\usepackage[normalem]{ulem}
\usepackage{enumitem}
\usepackage{placeins}

\usepackage{amsmath,amssymb}
\usepackage{graphics,graphicx}
\usepackage{dcolumn,bm}
\usepackage{psfrag}
\usepackage{enumerate}

\newcommand{\cu}{\affiliation{Department of Physics, University of Calcutta, 92 Acharya Prafulla Chandra Road, Kolkata 700009, West Bengal, India}}

\begin{document}

\bibliographystyle{apsrev4-1}


\title{Generalized BChS Model with Group Interactions: Shift in the Critical Point and Mean-Field Ising Universality}


\author{Amit Pradhan}
\email{prdhanamit84@gmail.com}
\cu



\date{\today}

\begin{abstract}
We introduce a generalized version of the Biswas–Chatterjee–Sen (BChS) model \cite{Biswas} with group interactions of size $q$, extending the original pairwise interaction dynamics. Within a mean-field framework, we derive an exact expression for the critical noise $p_c(q)$, showing that it increases monotonically with $q$ and approaches $1/2$ in the large-$q$ limit, consistent with a Gaussian approximation. Despite this shift in the phase boundary, the critical behavior remains unchanged across all $q$: the order parameter scales as $(p_c(q)-p)^{1/2}$ and the relaxation timescale diverges as $|p-p_c(q)|^{-1}$, identical to the original BChS model \cite{Biswas}. Finite-size scaling of the Binder cumulant, order parameter, and its fluctuations confirm that the system belongs to the mean-field Ising universality class for all $q$. Our results demonstrate that higher-order interactions modify the location of the transition without altering its universality class.
\end{abstract}


\maketitle


\section{Introduction}

The study of collective behavior in social systems has attracted significant attention within the framework of statistical physics. In particular, opinion dynamics models aim to understand how individual interactions can lead to consensus or coexistence of different opinions. Several models have been proposed over the years, and comprehensive reviews of this field can be found in Refs. \cite{Loreto,Galam,Sen,Perc,Jusup,Toscani}, including recent overviews of kinetic exchange approaches to socio-economic systems.

A widely studied class of models is based on kinetic exchange dynamics, where opinions evolve through pairwise interactions in which individuals partially adopt the opinions of others during interactions. This approach was introduced in the context of opinion formation by Lallouache et al. \cite{Mehdi}. It was subsequently generalized to include individual conviction and influencing ability through two parameters \cite{Parongama}. An exact solution of a discrete version of the model, exhibiting an active–absorbing phase transition and incorporating multi-agent interactions, was later obtained in Ref. \cite{Soumyajyoti}. A detailed overview of the developments in kinetic exchange models is given in Ref. \cite{Bikash,soumyajyoti}.

Building on this framework, the Biswas–Chatterjee–Sen (BChS) model was introduced in Ref. \cite{Biswas}. This model exhibits a noise-driven order–disorder phase transition, and its critical behavior has been analyzed within the same framework \cite{Biswas}. More generally, it has been shown that a broad class of kinetic exchange opinion models with noise exhibit mean field Ising universality \cite{Sudip}.

Subsequent works have explored various extensions of the BChS model, including studies in higher dimensions \cite{Arnab}, and modifications of the dynamical rules such as the inclusion of additional opinion states and stochastic switching mechanisms \cite{Kathakali,kathakali}. Further developments have examined dynamical features such as virtual walks and temporal evolution \cite{Katha,Surajit}, as well as interactions between multiple groups \cite{Suchecki}.

In addition, several studies have explored the behavior of the BChS model on complex networks and heterogeneous structures \cite{Nuno,Raquel,David,Lima,Oliveira,Sousa}, highlighting the robustness of the phase transition and its universality properties under different interaction topologies. However, a systematic understanding of how the size of interacting groups affects the location of the phase transition and its universality class is still lacking.

Most existing studies are based on pairwise interactions, where opinions are updated through interactions between two individuals at a time. In contrast, real social interactions often occur in groups, where multiple individuals interact simultaneously and influence each other collectively. Recent developments in the study of temporal and higher-order networks, together with recent studies of multi-agent wealth exchange models, have emphasized the importance of such group interactions, showing that collective interactions beyond pairs can qualitatively change the system dynamics \cite{Bianconi,Holme,Iacopini,Battiston,Suchismita,Benson}.

Motivated by these observations, we introduce a generalized version of the Biswas–Chatterjee–Sen (BChS) model in which interactions occur within groups of size $q$, rather than through pairwise exchanges. Within a mean-field framework, we analytically derive the phase boundary separating ordered and disordered states and show that increasing the group size enhances the stability of the ordered phase, leading to a systematic increase in the critical noise 
$p_c(q)$. In the limit of large-$q$, the critical point approaches $1/2$, consistent with a Gaussian approximation.

Despite the shift in the transition point, the critical behavior remains unchanged: the system continues to exhibit mean-field Ising universality, indicating that higher-order interactions modify the location of the transition without altering its universality class.

\section{Model and dynamics}

We consider a generalized version of the Biswas–Chatterjee–Sen (BChS) model in which interactions extend beyond pairwise exchanges to involve a group of $q$ neighbors. The parameter $q$ controls the size of the interacting group and varies in the range $1\leq q\leq N-1$, where $N$ is the system size.

 In the limiting case $q=1$, the model reduces to the original BChS model introduced in Ref. \cite{Biswas}. At the other extreme, the case $q=N-1$ corresponds to the fully connected (mean-field) limit, where each agent interacts with all other agents in the system. 
 
 Each agent $i$ carries a discrete opinion $O_i(t)\in \{-1,0,+1\}$. At each time step, an agent $i$ is selected at random, and a group of $q$ neighbors is chosen randomly from the population. The agent experiences an effective social influence from this group given by
\begin{equation}
    \eta_i = \text{sgn}\left(\mu_i\sum_{k=1}^{q}O_k\right).
\end{equation}
Here $\sum_{k=1}^{q}O_k$ represents the total opinion of the selected group, and $\eta_i$ is the resulting net influence acting on agent $i$. The interaction parameter $\mu_i$ determines the nature of the interaction and is defined as
\begin{equation}
\mu_i=
\begin{cases}
\;\;+1, & \text{with probability } 1-p,\\[4pt]
-\,1, & \text{with probability } p,
\end{cases}
\end{equation}
where $p$ is the disorder parameter. A positive value $\mu_i=+1$ corresponds to a reinforcing interaction, where the agent tends to align with the group influence, while $\mu_i=-1$ represents an opposing interaction, leading the agent to react against the group.

The opinion of the selected agent is updated according to
\begin{equation}
\label{update rule}
    O_i(t+1) = \text{sgn}[O_i(t) + \eta_i].
\end{equation}

\section{Mean-field theory and phase structure}

\subsection{Rate equations}

In this subsection, we derive the mean-field rate equations governing the evolution of the system.
Within the mean-field approximation, spatial correlations between agents are neglected, and the $q$ neighbors of a given agent are assumed to be chosen randomly from the population. 

The state of the system is  characterized by the fractions $f_+$,$f_-$, and $f_0$, corresponding to opinions $+1$, $-1$, and $0$, respectively, with the normalization condition $f_++f_-+f_0=1$.

Consider a randomly selected group of $q$ agents. Let $n_+$,$n_-$, and $n_0$ denote the number of agents in this group with opinion $+1$,$-1$, and $0$, respectively, such that
\begin{equation}
    n_++n_-+n_0 = q.
\end{equation}
Under the mean-field assumption, the probability of observing a given configuration ($n_+,n_-,n_0$) is given by the multinomial distribution
\begin{equation}
    \mathcal{P}(n_+,n_-,n_0) = \frac{q!}{n_+!n_-!n_0!}f_+^{n_+}f_-^{n_-}f_0^{n_0}.
\end{equation}
The total opinion of the group is 
\begin{equation}
\label{total opinion}
    S = \sum_{k=1}^{q} O_k = n_+-n_-.
\end{equation}
The effective social influence acting on the agent is $\eta_i=\text{sgn}(\mu_i S)$. Therefore, the probabilities of the three possible outcomes $\eta=+1,-1,0$ can be written as
\begin{equation}
\label{socia_influence_prob}
\begin{split}
   P(\eta=+1)= P_+ = (1-p)P(S>0) + pP(S<0),\\
   P(\eta=-1) = P_- = (1-p)P(S<0) + pP(S>0),\\
    P(\eta=0) = P_0 = P(S=0).
\end{split}
\end{equation}
Here,
\begin{equation}
\label{total_influence_prob}
    \begin{split}
        P(S>0) = \sum_{n_+>n_-} \mathcal{P}(n_+,n_-,n_0),\\
        P(S<0) = \sum_{n_+<n_-} \mathcal{P}(n_+,n_-,n_0),\\
        P(S=0) = \sum_{n_+=n_-} \mathcal{P}(n_+,n_-,n_0).
    \end{split}
\end{equation}
Using the update rule [Eq. (\ref{update rule})], one can enumerate all allowed transitions between the three opinion states. The transition rates are expressed in terms of $P_+$ and $P_-$ as follows :
\begin{equation}
    \begin{split}
       \omega_{+1\to 0} = P_-, \quad \omega_{+1 \to -1} = 0,\\
       \omega_{-1\to 0} = P_+, \quad \omega_{-1\to+1} = 0,\\
       \omega_{0\to +1} = P_+, \quad \omega_{0 \to -1} = P_-.
    \end{split}
\end{equation}
Using the above transition rates, the time evolution of the fractions $f_+$ and $f_-$ can be written as 
\begin{equation}
\begin{split}
    \frac{df_+}{dt} = -f_+P_-+f_0P_+,\\
    \frac{df_-}{dt} = -f_-P_+ +f_0P_-.
\end{split}
\end{equation}
To simplify the analysis, it is convenient to express the dynamics in terms of the order parameter $O=f_+-f_-$ and the activity $s=f_++f_-$. After straightforward algebra, one obtains
\begin{equation}
\label{rate equation for O}
 \frac{dO}{dt} = \left(1-\frac{s}{2}\right)(P_+-P_-)-\frac{O}{2}(P_++P_-),
 \end{equation}
 \begin{equation}
 \label{rate equation for s}
    \frac{ds}{dt} = \left(1-\frac{3s}{2}\right)(P_++P_-)+\frac{O}{2}(P_+-P_-).
\end{equation}

\subsection{Fixed points and phase boundary}

The stationary behavior of the system is governed by the fixed points of the coupled evolution equations for the order parameter $O$ [Eq. (\ref{rate equation for O})] and activity $s$ [Eq. \ref{rate equation for s})]. In this subsection, we first analyze the stability of the disordered  fixed point and determine the critical line $p_c(q)$, and then examine the existence of symmetry broken ordered solutions.

\begin{figure}[h]
\includegraphics[ width=\linewidth]{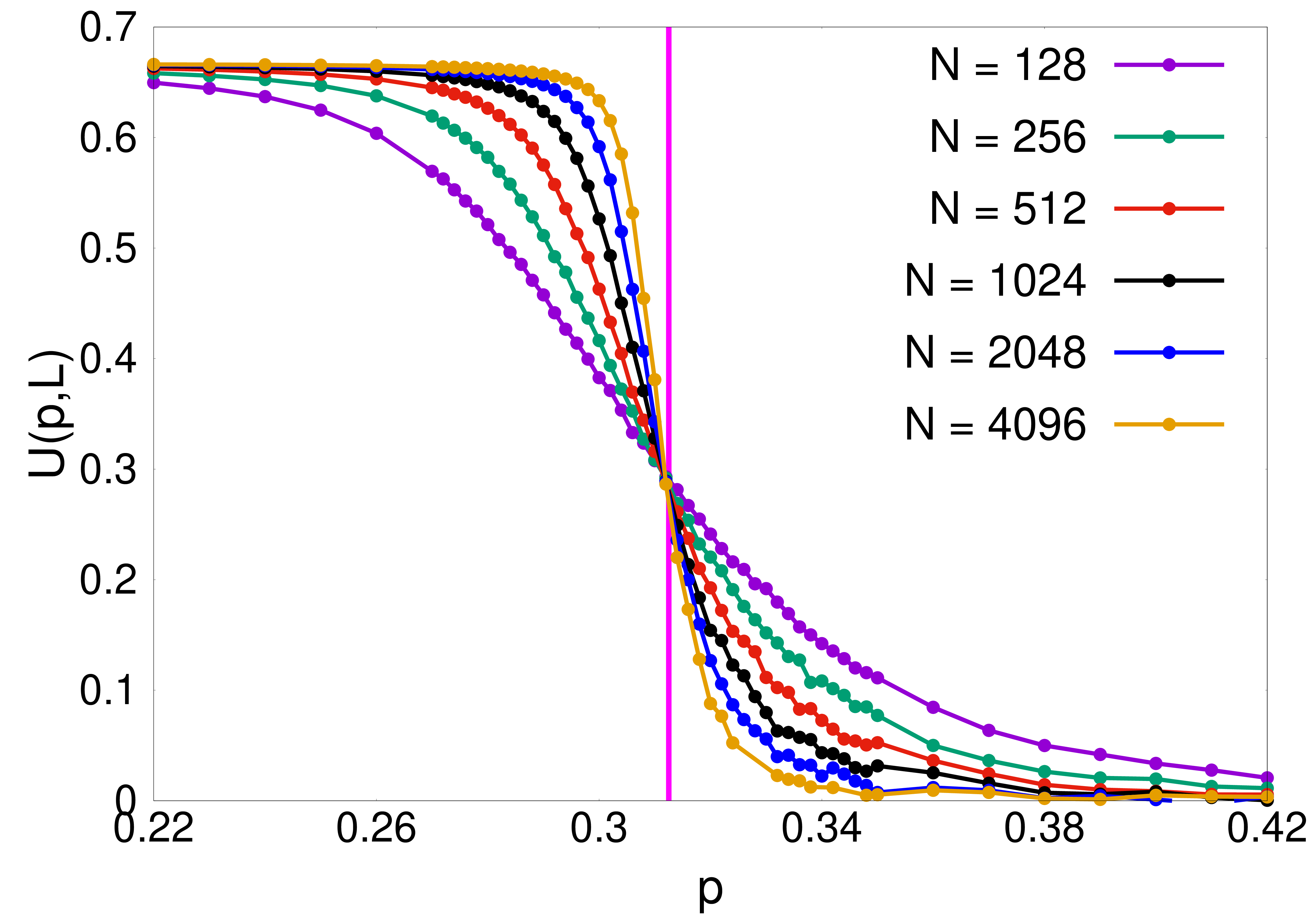} 
\caption{Binder cumulant $U(p,N)$ constructed from the moments of the order parameter $O$, as a function of $p$ is plotted for different system sizes $N$ with $q=2$. The curves intersect at a value consistent with $p_c(2)=5/16$, in agreement with the analytical prediction from Eq. (\ref{P_c for general q}).}
\label{BC_n2}
\end{figure}

\begin{description}
 \item[ Disordered fixed point and its stability :]
 The disordered fixed point can be obtained by setting the right-hand sides of Eqs. (\ref{rate equation for O}) and (\ref{rate equation for s}) to zero with $O=0$. Solving the resulting equation, one finds $s^*=2/3$, which immediately implies $f_+^*=f_-^*=f_0^*=1/3$. Therefore, the nontrivial disordered fixed point of the system is given by $(O^*, s^*) = (0, 2/3)$.

 To analyze the stability of this fixed point, we linearize the coupled evolution equations for $O$ and $s$ [Eq. (\ref{rate equation for O}),(\ref{rate equation for s})] around $(O^*,s^*)$. This leads to the Jacobian matrix evaluated at the fixed point. 
 
 A straightforward calculation shows that the Jacobian matrix is diagonal at $(O^*,s^*)$, with eigenvalues $\lambda_s=-1,\lambda_O$, corresponding to perturbations along the $s$ and $O$ directions, respectively.

 Since $\lambda_s<0$, deviations in the $s$ direction always decay. Therefore the stability of the disordered fixed point is completely determined by the sign of $\lambda_O$, which is obtained by linearizing the evolution equation of $O$ near this point.

 To do that, we begin by computing $P_+-P_-$ in the vicinity of $(O^*,s^*)=(0,2/3)$. Starting from the definitions used in Eq. (\ref{socia_influence_prob}) and (\ref{total_influence_prob}), we obtain
 \begin{equation}
 \begin{split}
     P_+-P_- = (1-2p)\bigg[\sum_{n_+>n_-}\mathcal{P}(n_+,n_-,n_0)\\ 
     - \sum_{n_+<n_-}\mathcal{P}(n_+,n_-,n_0)\bigg].
\end{split}
 \end{equation}
We combine the two sums into a single expression 
\begin{equation}
    P_+-P_- = (1-2p) \sum_{n_+,n_-} \text{sgn}(n_+-n_-)\mathcal{P}(n_+,n_-,n_0).
\end{equation}
Using $f_+ = (s+O)/2$, $f_-=(s-O)/2$ and $f_0=1-s$ we write
\begin{equation}
    f_+^{n_+}f_-^{n_-} = \left(\frac{s+O}{2}\right)^{n_+}\left(\frac{s-O}{2}\right)^{n_-}.
\end{equation}
Expanding this expression for small $O$, we obtain
\begin{equation}
   f_+^{n_+}f_-^{n_-} = \left(\frac{s}{2}\right)^{n_++n_-}\left[1+\frac{O}{s}(n_+-n_-)+\mathcal{O}(O^2)\right].
\end{equation}
Substituting this into the expression for $P_+-P_-$, the zeroth order contribution cancels due to symmetry, and the leading contribution becomes linear in $O$
\begin{equation}
    \begin{split}
        P_+-P_-= \frac{(1-2p)O}{s}\sum_{n_+,n_-}\text{sgn}(n_+-n_-)(n_+-n_-)\\
        \frac{q!}{n_+!n_-!n_0!} \left(\frac{s}{2}\right)^{n_++n_-}(1-s)^{n_0}.
    \end{split}
\end{equation}
Evaluating at the fixed point $s=2/3$,
\begin{equation}
    P_+-P_- = \frac{3}{2}(1-2p)O\sum_{n_+,n_-}\frac{q!}{n_+!n_-!n_0!}\left(\frac{1}{3}\right)^q|n_+-n_-|.
\end{equation}
We define 
\begin{equation}
\label{chi_q}
    \chi_q = \frac{3}{2}\sum_{n_+,n_-}\frac{q!}{n_+!n_-!n_0!}\left(\frac{1}{3}\right)^q|n_+-n_-|,
\end{equation}
so that
\begin{equation}
    P_+-P_-=(1-2p)\chi_qO.
\end{equation}
The quantity $\chi_q$ characterizes the average response of a group of size $q$ to a small perturbation in the order parameter.

We also use the identity 
\begin{equation}
    P_++P_- = 1-P_0,
\end{equation}
where
\begin{equation}
\label{P_0_disordered}
    P_0 = \sum_{n_+=n_-} \frac{q!}{n_+!n_-!n_0!}\left(\frac{1}{3}\right)^{q},
\end{equation}
is the probability that the total influence vanishes.

Substituting these results into the evolution equation of $O$ [Eq. (\ref{rate equation for O})] and evaluating at $s=2/3$, we obtain
\begin{equation}
    \frac{dO}{dt} = \left[\frac{2}{3}(1-2p)\chi_q-\frac{1}{2}(1-P_0)\right]O.
\end{equation}
Thus near the disordered fixed point,
\begin{equation}
\label{linearized_O_equation}
    \frac{dO}{dt} = \lambda_OO,
\end{equation}
where 
\begin{equation}
\label{lambda_O}
    \lambda_O = \frac{2}{3}(1-2p)\chi_q-\frac{1}{2}(1-P_0).
\end{equation}
The stability of the disordered fixed point is determined by the sign of $\lambda_O$. If $\lambda_O<0$, both eigenvalues ($\lambda_s$ and $\lambda_O$) are negative, and the disordered fixed point is stable. On the other hand, if $\lambda_O>0$, the fixed point becomes a saddle point, unstable along the $O$ direction. 

We will show later that this condition $\lambda_O>0$ is precisely the regime where two nontrivial ordered fixed points emerge and become stable.

The order-disorder transition point is obtained from the marginal condition $\lambda_O=0$, which yields
\begin{equation}
\label{P_c for general q}
    p_c(q) = \frac{1}{2}\left[1-\frac{3}{4}\frac{(1-P_0)}{\chi_q}\right].
\end{equation}
where $\chi_q$ and $P_0$ can be found from Eq. (\ref{chi_q}) and (\ref{P_0_disordered}) respectively.

Eq. (\ref{P_c for general q}) provides an explicit expression for the phase boundary $p_c(q)$, which constitutes the central analytical result of this work. We now turn to the existence of symmetry broken steady states characterized by a non zero order.

\item[Ordered fixed points :]
The ordered fixed points  $(\pm O^*,s^*)$ are obtained by setting the right hand sides of Eqs. (\ref{rate equation for O}) and (\ref{rate equation for s}) to zero with $O \neq 0$.

From Eq. (\ref{rate equation for O}), we obtain
\begin{equation}
\label{ordered_state_O^*}
    O^* = 2\left(1-\frac{s^*}{2}\right)\frac{P_+-P_-}{P_++P_-}.
\end{equation}
Substituting this into Eq. (\ref{rate equation for s}), the steady state condition for $s^*$ becomes
\begin{equation}
\label{ordered_state_s^*}
    \left(\frac{3s^*}{2}-1\right)(P_++P_-)^2 = \left(1-\frac{s^*}{2}\right)(P_+-P_-)^2.
\end{equation}
For general $q$, one has $P_++P_-=1-P_0$, while $P_+-P_-$ depends on the group statistics. Solving Eqs. (\ref{ordered_state_O^*})-(\ref{ordered_state_s^*}) self consistently determines the ordered state $(\pm O^*,s^*)$.

From these equations, one can eliminate $P_+-P_-$ to obtain
\begin{equation}
\label{general_expression_ordered_state}
    ({O^*})^2 = 4\left(1-\frac{s^*}{2}\right)\left(\frac{3s^*}{2}-1\right),
\end{equation}
which implies that a physical ordered solution exists only when $s^*>2/3$. To connect this condition with the stability of the disordered fixed point, we use Eq. (\ref{ordered_state_O^*}) to write
\begin{equation}
\label{expression_(1-2p)chi_q}
    (1-2p)\chi_q =\frac{1-P_0}{2\left(1-\frac{s^*}{2}\right)}.
\end{equation}
Substituting this into the expression of $\lambda_O$ in Eq. (\ref{lambda_O}), one obtains
\begin{equation}
  \lambda_O = (1-P_0)\frac{\left(\frac{3s^*}{2}-1\right)}{6\left(1-\frac{s^*}{2}\right)}.
\end{equation}
Since $1-P_0>0$ and $1-\frac{s^*}{2}>0$, it follows that
\begin{equation}
    s^*>\frac{2}{3} \quad \Leftrightarrow \quad \lambda_O>0,
\end{equation}
i.e., the ordered state exists only when $p<p_c(q)$, where the disordered fixed point $(0,2/3)$ becomes unstable (a saddle point).

For $q=1$, the transition probabilities simplify to
\begin{equation}
    P_+-P_- = (1-2p)O, \quad P_++P_- = s.
\end{equation}
Using these in Eq. (\ref{ordered_state_O^*}) and (\ref{ordered_state_s^*}), one obtains
\begin{equation}
    s^* = \frac{1-2p}{1-p}, \quad (O^*)^2 = \frac{1-4p}{(1-p)^2}.
\end{equation}
Thus, the existence condition for the ordered phase becomes
\begin{equation}
    \lambda_O = \frac{1-4p}{3}>0 \quad \Rightarrow \quad p<p_c = \frac{1}{4},
\end{equation}
in agreement with the original model \cite{Biswas}.
 \end{description}
 The general expression of the critical point obtained in Eq. (\ref{P_c for general q}) can now be verified for small values of $q$. For $q=1$, one finds $\chi_1 = 1$, $P_0=1/3$, which immediately gives $p_c(1)=1/4$, in agreement with the original model \cite{Biswas}. Similarly, for $q=2$, using $\chi_2=4/3,P_0=1/3$, we obtain $p_c(2)=5/16$. The Binder cumulant curves for different system sizes also intersect at a value consistent with $p_c(2) = 5/16$ (see Fig. \ref{BC_n2}), confirming the analytical prediction.

 Extending this to higher group sizes, for $q=3$ and $q=4$, using $P_0 = 7/27,19/81$ and $\chi_3=5/3,\chi_4 = 52/27$, one obtains $p_c(3) = 1/3\approx 0.33$ and $p_c(4)=73/208\approx 0.351$.

\begin{figure}[h]
\includegraphics[ width=\linewidth]{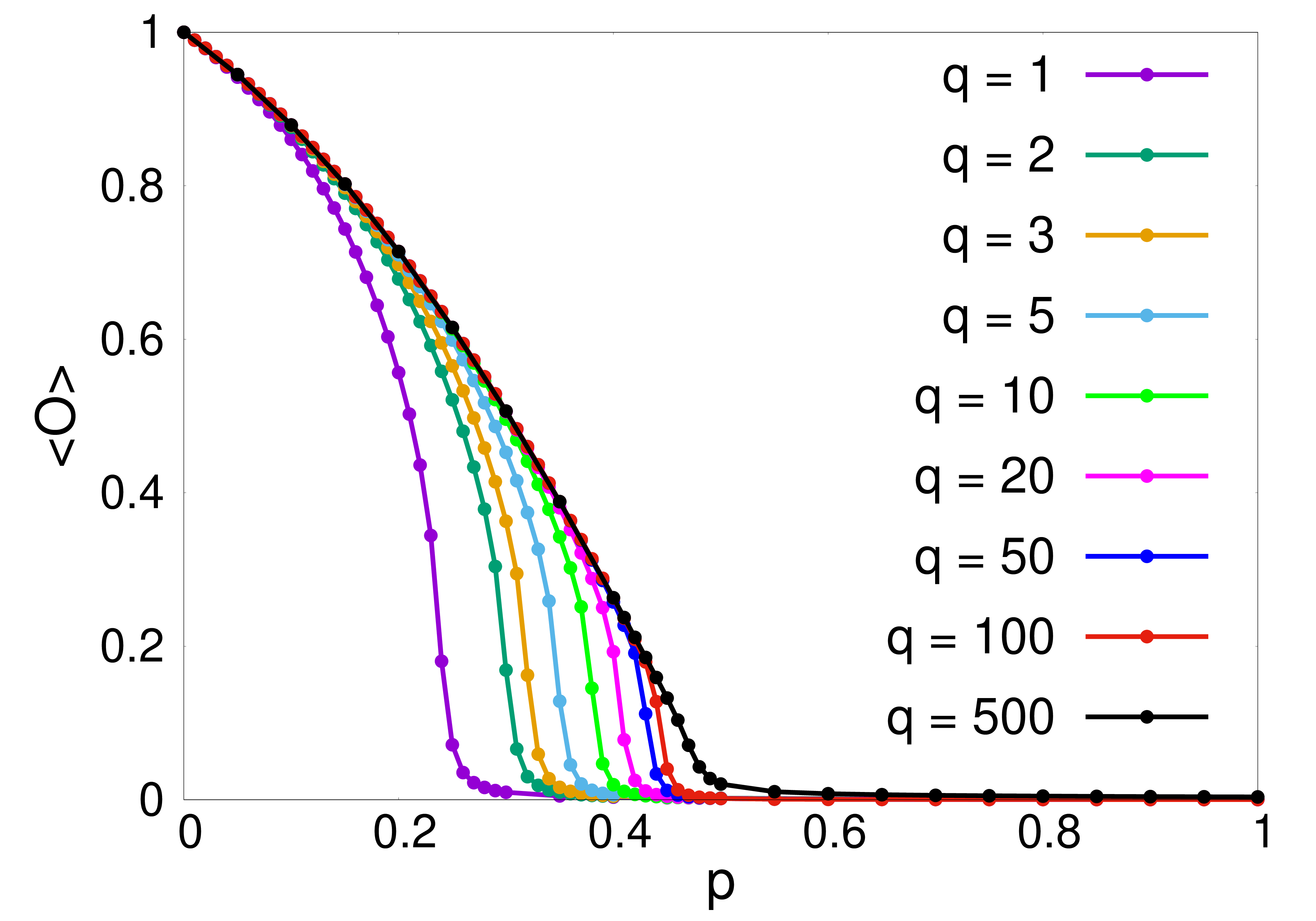} 
\caption{Ensemble averaged order parameter $\langle O\rangle$ as a function of the noise parameter $p$ is plotted for different group sizes $q$ with system size $N=1024$. The vanishing of $\langle O\rangle$ marks the transition point, which shifts to higher values of $p$ as $q$ increases, demonstrating that larger interaction groups stabilize the ordered phase. For large-$q$, the critical point approaches $p_c(q)\to 1/2$, and in particular, for $q=500$, we find $p_c(500)\approx 1/2$. }
\label{order_param_vs_p_q}
\end{figure}

 These results clearly show that the critical point increases with $q$, indicating that larger interaction groups enhance the stability of the ordered phase. Physically, increasing the group size reduces the sampling fluctuations in the local group opinion, which decreases as $1/\sqrt{q}$ and vanishes in the large-$q$ limit. As a result, the system can tolerate a larger amount of noise before losing order. This trend is also reflected in Fig. \ref{order_param_vs_p_q}, where the averaged order parameter $\langle O\rangle$ is plotted as a function of $p$ for different values of $q$, showing that the transition point shifts to higher values of $p$ as $q$
increases.

 \subsection{Large $\boldsymbol{q}$ limit : Gaussian Approximation}

 For large values of $q$, the exact expressions of $\chi_q$ [Eq. (\ref{chi_q})] and $P_0$ [Eq. (\ref{P_0_disordered})] involves sums over an exponentially large number of configurations and are therefore not convenient for analytical treatment. In this limit, a simpler description can be obtained using the central limit theorem.

 We note that the relevant stochastic quantity is the total opinion of the group $S$, defined in Eq. (\ref{total opinion}), which is a sum of independent and identically distributed random variables taking values $+1,-1$ and $0$ with probabilities $f_+,f_-$ and $f_0$, respectively.

 For large-$q$, $S$ approaches a Gaussian random variable with mean and variance given by
 \begin{equation}
     \langle S\rangle = qO, \quad \text{Var}(S) = qs.
 \end{equation}
 Thus the distribution of $S$ can be written as 
 \begin{equation}
     P(S) = \frac{1}{\sqrt{2\pi qs}} \exp\left[-\frac{(S-qO)^2}{2qs}\right].
 \end{equation}
 Using Eq. (\ref{socia_influence_prob}), one finds
 \begin{equation}
     P_+-P_- = (1-2p)[P(S>0)-P(S<0)].
 \end{equation}
 Evaluating this expression using the Gaussian distribution and introducing the scaled variable $x=(S-qO)/\sqrt{qs}$, we obtain
 \begin{equation}
     P(S>0)-P(S<0) = \text{erf}\left(\sqrt{\frac{q}{2s}}O\right).
 \end{equation}
 Therefore,
 \begin{equation}
 \label{P_+-P_-_CLT}
     P_+-P_- = (1-2p)\text{erf}\left(\sqrt{\frac{q}{2s}}O\right).
 \end{equation}
 To analyze the stability of the disordered fixed point $(O^*,s^*)=(0,2/3)$, we expand Eq. (\ref{P_+-P_-_CLT}) for small values of $O$. Using the leading order behavior of the error function, we obtain
 \begin{equation}
     P_+-P_- \approx (1-2p)\sqrt{\frac{2q}{\pi s}}O=(1-2p)\chi_qO,
 \end{equation}
 where the effective response function of the group
 \begin{equation}
     \chi_q = \sqrt{\frac{2q}{\pi s}} \quad \Rightarrow \quad  \chi_q = \sqrt{\frac{3q}{\pi}} \quad \text{at} \quad s^*=\frac{2}{3}.
 \end{equation}
 Similarly the probability of a neutral group is
 \begin{equation}
     P_0 = P(S=0) = \frac{1}{\sqrt{2\pi qs}} \quad \Rightarrow \quad P_0 =\sqrt{\frac{3}{4\pi q}} \quad \text{at} \quad s^*=\frac{2}{3}.
 \end{equation}
 Substituting these large-$q$ expressions into Eq. (\ref{P_c for general q}), we obtain the critical point for large-$q$ as
 \begin{equation}
 \label{p_c_large_q}
     p_c(q) = \frac{1}{2}\left[1-\frac{3}{4}\frac{\left(1-\sqrt{\frac{3}{4\pi q}}\right)}{\sqrt{\frac{3q}{\pi}}}\right].
 \end{equation}
 Thus,
 \begin{equation}
     p_c(q) \to \frac{1}{2} \quad \text{as} \quad q\to \infty.
 \end{equation}

\begin{figure}[h]
\includegraphics[ width=\linewidth]{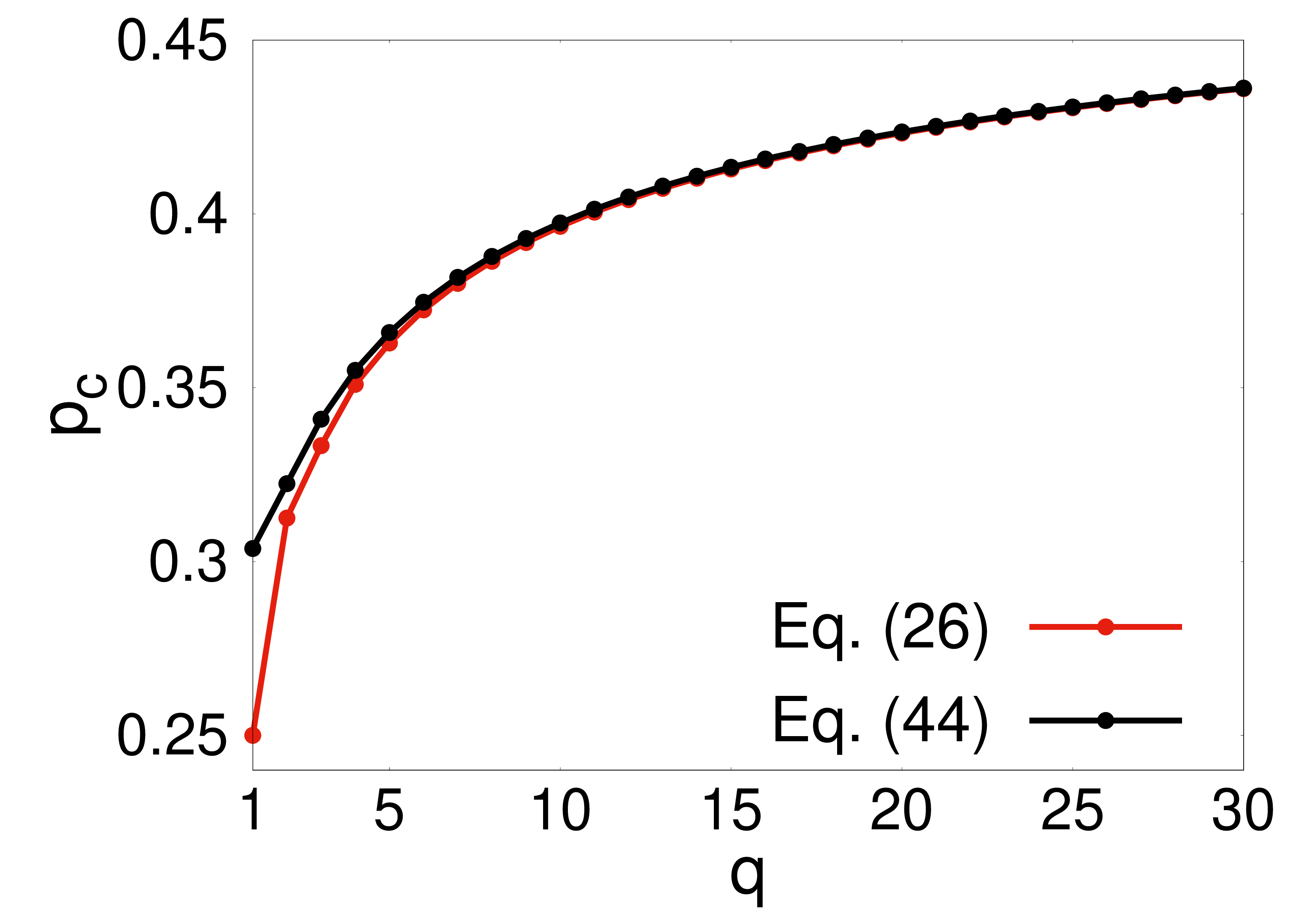} 
\caption{critical noise $p_c(q)$ is plotted as a function of group size $q$. Red line points correspond to the exact numerical evaluation based on Eq. (\ref{P_c for general q}), while black line points represent the asymptotic expression given by Eq. (\ref{p_c_large_q}). The two datasets converge rapidly with increasing $q$, becoming indistinguishable for $q\gtrsim 10$, confirming the validity of the Gaussian approximation in the large-$q$ limit.}
\label{critical_point_vs_q}
\end{figure}

 To validate the accuracy of the asymptotic expression in Eq. (\ref{p_c_large_q}), we numerically compute the critical point $p_c(q)$ using the exact mean-field formulation in Eq. (\ref{P_c for general q}) and compare it with the large-$q$ approximation. As shown in Fig. \ref{critical_point_vs_q}, the two results are in excellent agreement for sufficiently large-$q$, with an complete overlap observed for $q\gtrsim 10$. This demonstrates that the central limit (Gaussian) approximation provides an accurate description of the system behavior already at moderate values of $q$. The convergence of the two curves further confirms that $p_c(q)\to 1/2$ in the large-$q$ limit.

 This limiting value can also be understood physically. For large group sizes, fluctuations in the sampled group opinion become negligible, so the dynamics is governed only by the competition between the ordering tendency $1-p$ and the disordering tendency $p$. In the fully connected limit, the transition occurs when these two tendencies balance, giving $p_c=1/2$. For smaller $q$, fluctuations in the sampled group opinion weaken the effective ordering tendency, so the ordered phase is destroyed at lower values of noise.

 Finally, we emphasize that the expression in Eq. (\ref{p_c_large_q}) has been derived within a central limit (Gaussian) approximation and is therefore valid only for large values of $q$. For small $q$, the discrete nature of the opinion variables becomes important, and deviations from the asymptotic form arise, as the Gaussian approximation no longer captures the underlying discrete fluctuations accurately.

\vspace{5mm}

 \section{Critical Behavior and Universality }

 To characterize the nature of the phase transition, we analyze the critical behavior of the model within the mean-field framework. In particular, we derive the scaling behavior of the order parameter near the critical point $p=p_c(q)$ to extract the associated critical exponent $\beta$. We also examine the relaxation dynamics in the vicinity of the transition to determine the divergence of the characteristic timescale and obtain the exponent $\nu z$. These results allow us to assess the universality class of the model and compare it with that of the 
 original BChS dynamics.

 \subsection{Order parameter scaling near $p_c(q)$}

 To determine the behavior of the order parameter near the transition, we analyze the ordered solution in the vicinity of the critical point $p\to p_c(q)$. Close to the transition, the ordered fixed point satisfies 
 \begin{equation}
     s^*=\frac{2}{3}+\delta, \quad \delta \ll1.
 \end{equation}
 Using Eq. (\ref{general_expression_ordered_state}) and expanding for small $\delta$, we obtain to leading order
 \begin{equation}
     (O^*)^2 \approx 4\delta \quad \Rightarrow \quad O^* \sim \sqrt{\delta}.
 \end{equation}
 To relate $\delta$ with the distance from criticality, we use  Eq. (\ref{expression_(1-2p)chi_q}). Expanding around $s^*=2/3$ and comparing with its value at the critical point, one finds that
 \begin{equation}
     \delta = \frac{32}{9}\frac{\chi_q}{1-P_0}(p_c-p) \quad \Rightarrow \quad \delta \propto (p_c-p).
 \end{equation}
 Combining the above results, we obtain the scaling behavior of the order parameter near the transition 
 \begin{equation}
     O^* \sim (p_c-p)^{\frac{1}{2}},
 \end{equation}
 independent of $q$, indicating mean-field Ising criticality, identical to that of the original BChS model \cite{Biswas}.

 \subsection{Divergence of the relaxation timescale}

 To quantify how fast the system returns to the disordered fixed point, we consider a small deviation of the order parameter from $O^*=0$, i.e.,
 \begin{equation}
     O(t) = \epsilon(t), \quad \epsilon\ll1.
 \end{equation}
 From Eq. (\ref{linearized_O_equation}), one obtains 
 \begin{equation}
     \frac{d\epsilon}{dt} = \lambda_O\epsilon,
 \end{equation}
 where $\lambda_O<0$ in the disordered phase. The solution is $\epsilon(t)=\epsilon(0)e^{\lambda_O t}$, which defines the relaxation timescale as 
 \begin{equation}
     \tau = \frac{1}{|\lambda_O|},
 \end{equation}
  where $\lambda_O$ is defined in Eq. (\ref{lambda_O}).

  Near the critical point $p_c(q)$, expanding $\lambda_O$ to leading order in $p-p_c(q)$ yields
  \begin{equation}
      \lambda_O\approx -\frac{4}{3}\chi_q(p-p_c(q))  \sim (p-p_c(q)).
  \end{equation}
  As a result, the relaxation timescale diverges as 
  \begin{equation}
      \tau\sim |p-p_c(q)|^{-1}.
  \end{equation}
  Thus the associated critical exponent is $\nu z=1$, independent of $q$, indicating mean-field Ising universality of the dynamical scaling, consistent with the behavior of the original BChS model \cite{Biswas}. These analytical predictions for the critical exponents will be tested numerically using finite-size scaling in Sec. V.

  \section{ Numerical estimation of critical exponents using Finite-size scaling}

  To verify the analytical predictions for the critical behavior, we perform a finite-size scaling (FSS) analysis of the model. In Sec. IV, we obtained the mean-field critical exponents $\beta=1/2$ and $\nu z=1$. Here, we independently estimate the static critical exponents $\beta$, $\nu$, and $\gamma$ from numerical simulations using standard FSS techniques.

  The Binder cumulant is defined as 
  \begin{equation}
      U(p,N) = 1- \frac{\langle O^4\rangle}{3\langle O^2\rangle^2},
  \end{equation}
 and the susceptibility (fluctuation of the order parameter) as
 \begin{equation}
     \chi_O = N\left(\langle O^2\rangle-\langle O\rangle^2\right),
 \end{equation}
 where $O$ denotes the order parameter and $\langle.\rangle$ represents averaging over different realizations.

 Near the critical point $p_c(q)$, these quantities are expected to obey the standard FSS forms:
 \begin{equation}
 \begin{split}
     U(p,N) = \mathcal{F}\left[(p-p_c(q))N^{\frac{1}{\nu}}\right],\\ \chi_O(p,N) = N^{\frac{\gamma}{\nu}}\mathcal{G}\left[(p-p_c(q))N^{\frac{1}{\nu}}\right],\\ \langle O\rangle(p,N) = N^{-\frac{\beta}{\nu}}\mathcal{H}\left[(p-p_c(q))N^{\frac{1}{\nu}}\right].
 \end{split}
 \end{equation}

\begin{figure}[!htbp]
\includegraphics[ width=\linewidth]{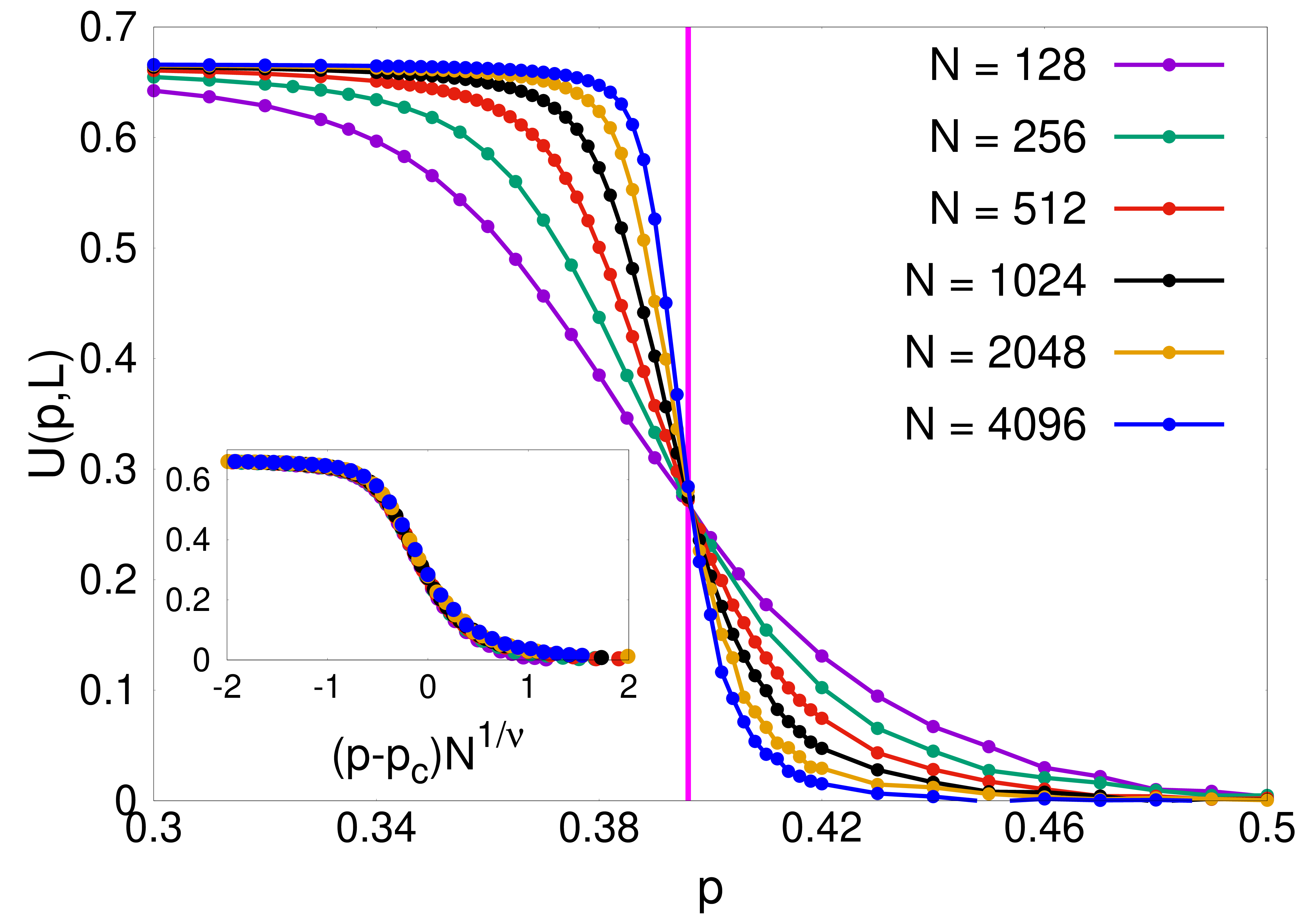} 
\caption{Binder cumulant $U(p,N)$ is plotted as a function of $p$ for different sizes $N$ at $q=10$. The curves intersect at $p_c(10)\approx 0.396$. Inset shows the data collapse of $U(p,N)$ as a function of $(p-p_c)N^{1/\nu}$ with $1/\nu = 1/2$.}
\label{BC_n10}
\end{figure}

 Fig. \ref{BC_n10} shows the Binder cumulant $U(p,N)$ as a function of $p$ for different system sizes for $q=10$. The curves intersect at $p_c(10)\approx 0.396$, and a good data collapse is obtained using $1/\nu = 1/2$, indicating $\nu = 2$.

\begin{figure}[!htbp]
\includegraphics[ width=\linewidth]{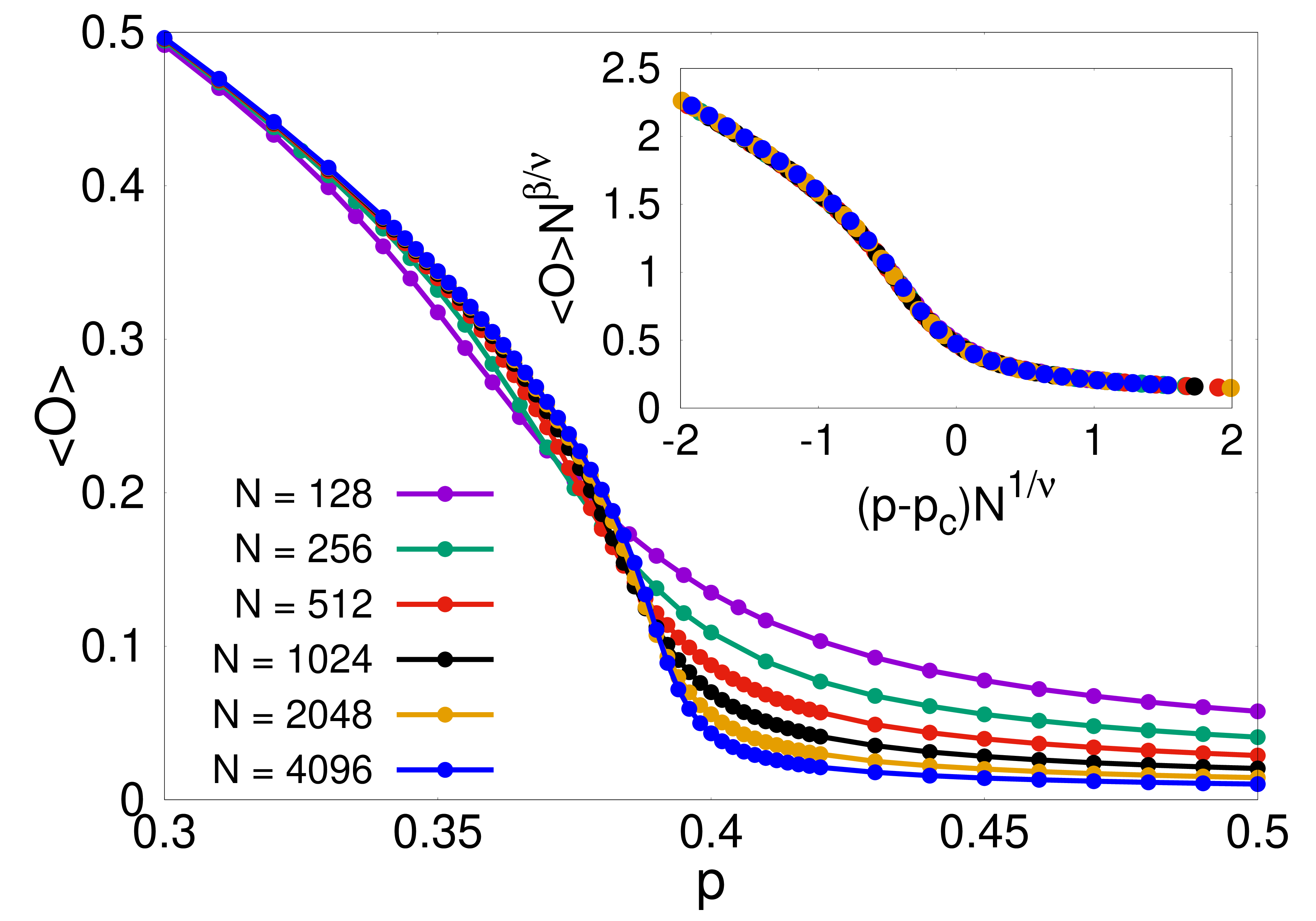} 
\caption{Steady state order parameter $\langle O\rangle$ is plotted as a function of $p$ for different system sizes $N$ at $q=10$. Inset shows the data collapse of $\langle O\rangle N^{\beta/\nu}$ versus $(p-p_c)N^{1/\nu}$ with $\beta/\nu =1/4$ and $1/\nu = 1/2$.}
\label{order_param_n10}
\end{figure}

 In Fig. \ref{order_param_n10}, we present the steady state order parameter $\langle O\rangle$. The data collapse is achieved for $\beta/\nu = 1/4$ and $1/\nu = 1/2$, yielding $\beta=1/2$.

\begin{figure}[!htbp]
\includegraphics[ width=\linewidth]{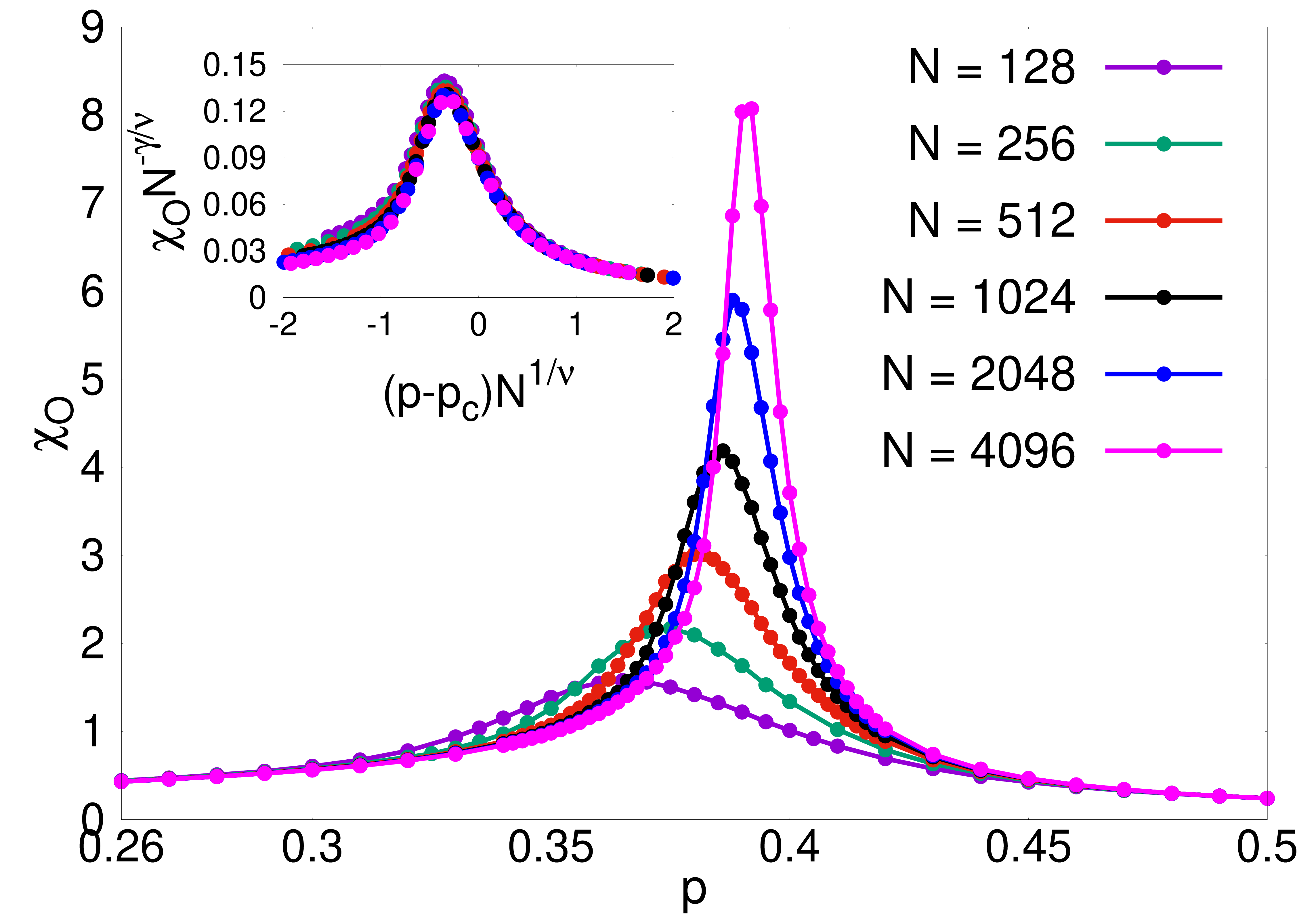} 
\caption{Susceptibility $\chi_O$ is plotted as a function of $p$ for different system sizes $N$ at $q=10$. Inset shows the data collapse of $\chi_ON^{-\gamma/\nu}$ versus $(p-p_c)N^{1/\nu}$ with $\gamma/\nu =1/2$ and $1/\nu = 1/2$.}
\label{fluctuation_n10}
\end{figure}

 Finally, Fig. \ref{fluctuation_n10} shows the susceptibility $\chi_O$, which collapses for $\gamma/\nu = 1/2$ and $1/\nu = 1/2$, giving $\gamma=1$.

 These results confirm that the critical exponents $\beta=1/2,\nu=2,$ and $\gamma=1$ are independent of $q$ and consistent with the mean-field Ising universality class. Although the plots are shown here for $q=10$, identical scaling is observed for other values of $q$.

We conclude with the remark that the estimated values of the critical exponents $\beta$ and $\gamma$ are consistent with the mean-field Ising values. Interpreting $\nu$ as $\nu^\prime d$, where $d$ is the effective dimension in this long range interacting system, and taking $d=4$ as in the original model \cite{Biswas} and in small-world-like networks \cite{Hong,Chatterjee}, we obtain $\nu=\nu^\prime d=2$ (with $\nu^\prime=1/2$), consistent with the mean-field Ising universality class.


\section{Summary and Conclusion}

 In this work, we have introduced and analyzed a generalized version of the Biswas–Chatterjee–Sen (BChS) model incorporating group interactions of size $q$. Within a mean-field framework, we derived the phase structure of the model and obtained an explicit expression for the critical point $p_c(q)$, demonstrating its systematic dependence on the group size. In particular, we showed that $p_c(q)\to 1/2$ in the large-$q$ limit, consistent with a central limit (Gaussian) approximation.

We further investigated the critical behavior of the model and found that the order parameter exhibits mean-field scaling with exponent $\beta=1/2$, while the relaxation dynamics is characterized by $\nu z=1$. These analytical predictions were tested through numerical simulations using finite-size scaling. From the scaling of the Binder cumulant, order parameter, and susceptibility, we independently estimated the critical exponents $\beta$, $\nu$, and $\gamma$, and found them to be consistent with the mean-field Ising universality class, as observed in the original model \cite{Biswas}.

Our results highlight two main findings. First, increasing the interaction group size shifts the transition point to higher values of noise, showing that larger groups make the ordered phase more stable. In other words, collective order becomes more robust when agents interact in larger groups. Second, although the location of the transition changes with $q$, the critical behavior remains unchanged for all group sizes. This shows that while higher-order interactions affect the stability of the ordered state, they do not alter the universal scaling behavior of the system. The persistence of mean-field Ising universality for all $q$ reflects the mean-field nature of the model, indicating that the interaction structure primarily affects the phase boundary rather than the critical scaling. 

The importance of group interactions in opinion dynamics has been recognized in several earlier studies. Such interactions describe situations in which an individual's opinion is influenced by several people acting together rather than by isolated pairwise interactions. The original $q$-voter model introduced by Castellano et al. \cite{Castellano} replaced pairwise interactions in the voter model with group interactions and showed that this simple modification can produce qualitatively distinct collective behavior and different routes to consensus. Later studies further showed that various forms of group influence and changes in the interaction group sizes can significantly affect the dynamics of opinion formation and collective decision making \cite{Pratik,Doniec}. Our generalized BChS model extends this line of research by explicitly incorporating group interactions into a kinetic exchange opinion model. We find that increasing the interaction group size systematically shifts the phase boundary, making the ordered phase more stable, while the universal critical behavior remains unchanged. Thus, in the present model, group interactions do not change the universality class but do modify the conditions under which collective order emerges. This observation is also relevant to the recent discussion on the role of higher-order interactions in complex systems \cite{Peixoto}. Although our work does not address the general question of how higher-order interactions should be represented, it demonstrates that explicit group interactions provide a simple and physically meaningful framework for understanding how interaction size influences collective dynamics.

\begin{acknowledgments}
AP would like to acknowledge University Grant Commission (UGC), Govt. of India for financial support (NTA REFERENCE. NO.: 241610061476). AP also thanks Parongama Sen for useful discussions and for a critical reading of the manuscript.
\end{acknowledgments}


\begin{thebibliography}{99}
\bibitem{Loreto}
C. Castellano, S. Fortunato, and V. Loreto, {\em Statistical physics of social dynamics},
Rev. Mod. Phys. {\bf 81}, 591-646  (2009).

\bibitem{Galam}
S. Galam, Sociophysics, {\em A Physicist’s Modeling of Psycho-political Phenomena}, Springer
New York, NY (2016).

\bibitem{Sen}
P. Sen and B. K. Chakrabarti, {\em Sociophysics: An introduction}, Oxford University
Press, Oxford (2014).

\bibitem{Perc}
M. Perc, {\em The social physics collective}, Scientific Reports {\bf 9}, 16549 (2019).

\bibitem{Jusup}
M. Jusup et al.,{\em Social physics}, Physics Reports {\bf 948}, 1–148 (2022).

\bibitem{Toscani}
G. Toscani, P. Sen, S. Biswas, {\em Kinetic exchange models of societies and economies}, Philos Trans A {\bf 380}, 20210170 (2022).

\bibitem{Mehdi}
M. Lallouache, A. S. Chakrabarti, A. Chakraborti, and B. K. Chakrabarti, {\em Opinion formation in kinetic exchange models: Spontaneous symmetry-breaking transition}, Phys. Rev. E {\bf 82}, 056112 (2010).

\bibitem{Parongama}
P. Sen, {\em Phase transitions in a two-parameter model of opinion dynamics with random kinetic exchanges}, Phys. Rev. E {\bf 83}, 016108  (2011).

\bibitem{Soumyajyoti}
S. Biswas, {\em Mean-field solutions of kinetic-exchange opinion models}, Phys. Rev. E {\bf 84}, 056106 (2011).

\bibitem{Bikash}
 S. Biswas, A. Chatterjee, P. Sen, S. Mukherjee  and B. K. Chakrabarti, {\em Social dynamics through kinetic exchange: the BChS model}, Front. Phys. 11, 1196745 (2023).

\bibitem{soumyajyoti}
S. Biswas, A. K. Chandra, A. Chatterjee, and B. K Chakrabarti, {\em Phase transitions and non-equilibrium relaxation in kinetic models of opinion formation}, Journal of Physics: Conference Series {\bf 297}, 012004 (2011).

\bibitem{Biswas}
S. Biswas, A. Chatterjee, and P. Sen, {\em Disorder induced phase transition in kinetic models of opinion dynamics}, Physica A: Statistical Mechanics and its Applications {\bf 391}, 3257–3265 (2012).

\bibitem{Sudip}
S. Mukherjee, S. Biswas, A. Chatterjee, and B. K. Chakrabarti, {\em The Ising universality class of kinetic exchange models of opinion dynamics}, Physica A: Statistical Mechanics and its Applications {\bf 567}, 125692 (2021).

\bibitem{Arnab}
S. Mukherjee and A. Chatterjee, {\em Disorder-induced phase transition in an opinion dynamics model: Results in two and three dimensions}, Phys. Rev. E {\bf 94}, 062317 (2016).

\bibitem{Kathakali}
K. Biswas and P. Sen, {\em Nonequilibrium dynamics in a three-state opinion-formation model with stochastic extreme switches}, Phys. Rev. E {\bf 106}, 054311  (2022).

\bibitem{kathakali}
K. Biswas and P. Sen, {\em Opinion formation models with extreme switches and disorder: critical behavior and dynamics}, Phys. Rev. E {\bf 107}, 054106  (2023).

\bibitem{Katha}
K. Biswas and P. Sen, {\em Virtual walks and phase transitions in the two-dimensional Biswas-Chatterjee–Sen model with extreme switches}, Phys. Rev. E {\bf 110}, 024105 (2024).

\bibitem{Surajit}
S. Saha and P. Sen, {\em Virtual walks inspired by a mean-field kinetic exchange model of opinion dynamics}, Philos Trans A {\bf 380}, 20210168 (2022).

\bibitem{Suchecki}
K. Suchecki, K. Biswas, J. A. Holyst, and P. Sen, {\em Biswas-Chatterjee-Sen kinetic exchange opinion model for two connected groups}, Phys. Rev. E {\bf 112}, 014304 (2025).

\bibitem{Nuno}
N. Crokidakis, {\em Noise and disorder: Phase transitions and universality in a model of opinion formation}, International Journal of Modern Physics C {\bf 27}, 1650060 (2016).

\bibitem{Raquel}
M. T. S. A. Raquel et al., {\em Non-equilibrium kinetic Biswas–Chatterjee–Sen model on complex networks}, Physica A {\bf 603}, 127825 (2022).

\bibitem{David}
David S. M. Alencar wt al., {\em Opinion Dynamics Systems on Barabási–Albert Networks: Biswas–Chatterjee–Sen Model}, Entropy {\bf 25}(2), 183 (2023).

\bibitem{Lima}
 F. W. S. Lima, M. A. Sumour, A. A. Moreira, and A. D. Araujo, {\em Non-equilibrium BCS model on Apollonian networks}, Physica A: Statistical Mechanics and its Applications {\bf 571}, 125834 (2021).

 \bibitem{Oliveira}
 G. S. Oliveira et al., {\em Biswas–Chatterjee–Sen Model Defined on Solomon Networks in $1\leq d\leq 6$-Dimensional Lattices}, Entropy {\bf 27}(3), 300 (2025).

 \bibitem{Sousa}
 G. S. Oliveira, T. A. Alves, G. A. Alves, F. W. Lima, and J. A. Plascak, {\em Biswas–Chatterjee–Sen model on Solomon networks with two three-dimensional lattices}, Entropy {\bf 26}(7), 587 (2024).

 \bibitem{Bianconi}
 G. Bianconi, {\em Higher-Order Networks}, Cambridge University Press, Cambridge (2021).

\bibitem{Holme}
P. Holme and J. Saramäki, {\em Temporal networks}, Physics Reports {\bf 519}, 97–125 (2012).

\bibitem{Iacopini}
I. Iacopini, G. Petri, A. Barrat, and V. Latora, {\em Simplicial models of social contagion}, Nature Communications {\bf 10}, 2485 (2019).

\bibitem{Battiston}
F. Battiston et al., {\em Networks beyond pairwise interactions: Structure and dynamics}, Physics Reports {\bf 874}, 1–92 (2020).

\bibitem{Suchismita}
S. Banerjee, {\em Exploring the impact of multi-agent wealth exchange model on inequality reduction}, Physica A: Statistical Mechanics and its Applications {\bf 693}, 131553 (2026).

\bibitem{Benson}
A. R. Benson, D. F. Gleich, J. Leskovec, {\em Higher-order organization of complex networks}, Science {\bf 353}, 163–166 (2016).

\bibitem{Hong}
H. Hong, B. J. Kim, M. Y. Choi, {\em Comment on “Ising model on a small world network”}, Phys. Rev. E {\bf 66}, 018101 (2002).

\bibitem{Chatterjee}
A. Chatterjee, P. Sen, {\em Phase transitions in an Ising model on a Euclidean network}, Phys. Rev. E {\bf 74}, 036109 (2006).

\bibitem{Castellano}
C. Castellano, M. A. Munoz, R. Pastor-Satorras, {\em Nonlinear q-voter model}, Phys. Rev. E {\bf 80}, 041129 (2009).

\bibitem{Pratik}
P. Mullick and P. Sen, {\em Social influence and consensus building: Introducing a q-voter model with weighted influence}, PLoS One {\bf 20}, e0316889 (2025).

\bibitem{Doniec}
M. Doniec, P. Mullick, P. Sen, K. Sznajd-Weron, {\em Modeling biases in binary decision-making within the generalized nonlinear q-voter model}, Chaos 35, 043133 (2025).

\bibitem{Peixoto}
T. P. Peixoto, L. Peel, T. Gross, M. De Domenico, {\em Graphs are maximally expressive for higher-order interactions}, arXiv:2602.16937 (2026).
\end{thebibliography}
\end{document}